\documentclass[fleqn,usenatbib]{mnras}
\DeclareRobustCommand{\VAN}[3]{#2}
\let\VANthebibliography\thebibliography
\def\thebibliography{\DeclareRobustCommand{\VAN}[3]{##3}\VANthebibliography}

\usepackage{graphicx}	
\usepackage{amsmath}	
\usepackage{amssymb}	

\usepackage{url}

\title[Galaxy rotation in JADES]{The distribution of galaxy rotation in JWST Advanced Deep Extragalactic Survey}

\author[Lior Shamir]{
Lior Shamir,$^{1}$\thanks{E-mail: lshamir@mtu.edu}
\\
$^{1}$Kansas State University, Manhattan, KS, 66506, USA\\
}

\date{Accepted xxxx xx xx. Received xxxx xx xx; in original form xxxx xx xx}

\pubyear{2025}

\begin{document}
\label{firstpage}
\pagerange{\pageref{firstpage}--\pageref{lastpage}}
\maketitle

\begin{abstract}

JWST provides a view of the Universe never seen before, and specifically fine details of galaxies in deep space. JWST Advanced Deep Extragalactic Survey  (JADES) is a deep field survey, providing unprecedentedly detailed view of galaxies in the early Universe. The field is also in relatively close proximity to the Galactic pole. Analysis of spiral galaxies by their direction of rotation in JADES shows that the number of galaxies in that field that rotate in the opposite direction relative to the Milky Way galaxy is $\sim$50\% higher than the number of galaxies that rotate in the same direction relative to the Milky Way. The analysis is done using a computer-aided quantitative method, but the difference is so extreme that it can be noticed and inspected even by the unaided human eye. These observations are in excellent agreement with deep fields taken at around the same footprint by HST and JWST. The reason for the difference may be related to the structure of the early Universe, but it can also be related to the physics of galaxy rotation and the internal structure of galaxies. In that case the observation can provide possible explanations to other puzzling anomalies such as the $H_o$ tension and the observation of massive mature galaxies at very high redshifts.



\end{abstract}




\section{Introduction}
\label{introduction}

James Webb Space Telescope (JWST) has introduced unprecedented imaging power, allowing it to capture high visual details of astronomical objects in deep space. The ability to identify shapes of objects in the very early Universe has a transformative impact on astronomy and cosmology. An example is the galaxies identified at very high redshifts \citep{adams2023discovery,boylan2023stress,bradley2023high,carniani2024spectroscopic}, such as JADES-GS-z14-0 at redshift of $\sim$14.2, and just $\sim$0.25 Gyr after the Big Bang \citep{schouws2024detection,carniani2024spectroscopic,jones2024jades,helton2024jwst}. Galaxies at unexpectedly high redshift also include Milky Way-like spiral galaxies \citep{costantin2023milky,jain2024grand}, showing that such galaxies are also present at relatively high redshift ranges \citep{kuhn2024jwst}. Although spiral galaxies at unexpectedly high redshifts were known before JWST was launched \citep{tsukui2021spiral}, the visual observations enabled by JWST are considered surprising given the current cosmological and galaxy formation theories \citep{forconi2023early,melia2023cosmic,adil2023dark,gupta2023jwst,xiao2023massive,boylan2023stress}. 

Additionally, the yet unexplained $H_o$ tension \citep{wu2017sample,mortsell2018does,bolejko2018emerging,davis2019can,pandey2020model,camarena2020local,di2021realm,riess2022comprehensive} introduces a substantial challenge to cosmology, and it has been suggested that the $H_o$ tension and high-redshift galaxies observed by JWST are linked \citep{shen2024early}. While research is bound to continue, the unexpected observations made so far by JWST have been argued to be in tension with standard cosmology  
\citep{gupta2023jwst,lovell2023extreme,dolgov2023tension,wang2023jwst,munoz2024reionization,forconi2023early,gupta2024dark,gupta2401testing}.

One of the observations enabled by the ability of JWST to identify high visual details of galaxies is the alignment between the galaxy direction of rotation as observed by JWST and the direction of rotation of the Milky Way \citep{Shamir_2024}. Namely, JWST shows a much higher number of galaxies that rotate in the opposite direction relative to the Milky Way. That can be observed in JWST deep fields taken at close proximity to the Galactic pole. When spiral galaxies are located at around the Galactic pole, their direction of rotation can determine whether they rotate in the same direction relative to the Milky Way, or in the opposite direction relative to the Milky Way \citep{Shamir_2024}.

\begin{figure*}
\centering
\includegraphics[scale=0.17]{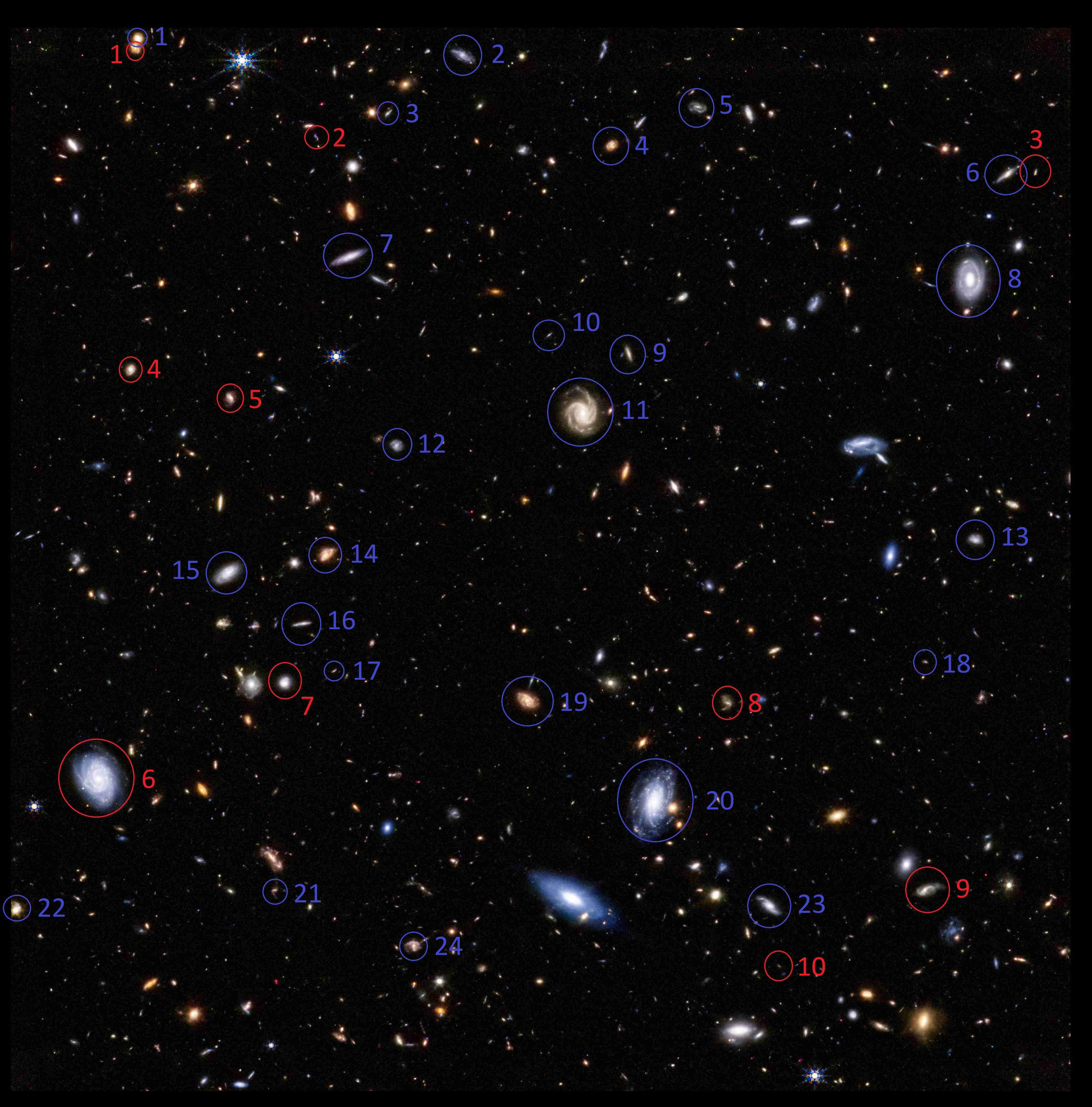}
\caption{Spiral galaxies imaged by JWST that rotate in the same direction relative to the Milky Way (red) and in the opposite direction relative to the Milky Way (blue). The number of galaxies rotating in the opposite direction relative to the Milky Way as observed from Earth is far higher \citep{Shamir_2024}.}
\label{jwst_small2}
\end{figure*}

A first observation of the higher prevalence of galaxies that rotate in opposite direction relative to the Milky Way in JWST deep fields was reported in \citep{Shamir_2024}. The analysis was based on a preliminary JWST deep field image taken inside the field of the Hubble Space Telescope (HST) Ultra Deep Field (UDF). The deep field was imaged in October 2022, and the image was released to the public on April 2023. Analysis of the field \citep{Shamir_2024} identified 33 galaxies with identifiable direction of rotation, where 23 of them rotated in the opposite direction relative to the Milky Way (p$\simeq$0.012). Figure~\ref{jwst_small2} shows the deep field annotated by the direction of rotation of the galaxies \citep{Shamir_2024}.

When done manually, the determination of the direction of rotation of a galaxy can be a subjective task, as different annotators might have different opinions regarding the direction towards a galaxy rotates. A simple example is the crowdsourcing annotation through {\it Galaxy Zoo 1} \citep{land2008galaxy}, where in the vast majority of the galaxies different annotators provided conflicting annotations. Therefore, the annotations shown in Figure~\ref{jwst_small2} were made by a computer analysis that followed a defined symmetric model \citep{Shamir_2024}. Yet, the advantage of the analysis of the relatively small JWST deep field is that it can be inspected by the human eye to ensure that the annotations of the galaxies are consistent, and that no population of non-annotated galaxies that could change the outcome of the analysis exists \citep{Shamir_2024}.

The difference between the number of galaxies that rotate in opposite directions was also noticed when using Earth-based telescopes \citep{macgillivray1985anisotropy,longo2011detection,shamir2012handedness,shamir2016asymmetry,shamir2019large,shamir2020patterns,shamir2020large,shamir2020pasa,shamir2021particles,shamir2021large,shamir2022new,shamir2022large,shamir2022asymmetry,shamir2022analysis}. Namely, it has been shown that the difference between the number of galaxies that rotate in opposite direction increases as the redshift gets higher \citep{shamir2019large,shamir2020patterns,shamir2022large,sym16101389}, which might suggest that the difference becomes larger in the deep Universe as imaged by JWST. On the other hand, several studies argued that the distribution is random \citep{iye1991catalog,land2008galaxy,hayes2017nature,tadaki2020spin,iye2020spin,petal}. These studies will be discussed in Section~\ref{other_reports} of this paper. But with the imaging power of JWST, the uneven distribution becomes clear, and can be verified even with the unaided human eye \citep{Shamir_2024}. This paper analyzes the distribution of spiral galaxies in the JADES survey. Any anomaly in the distribution can be related to the structure of the early universe, but can also be driven by the mysterious physics of galaxy rotation.

While several analyses using different instruments were performed, JWST introduces new opportunities to study the asymmetry in the early Universe. The imaging power of JWST is particularly meaningful because the magnitude of the asymmetry has been identified to grow as the redshift gets larger \citep{shamir2019large,shamir2020patterns,shamir2022large,sym16101389}, and therefore studying the asymmetry in deep fields can lead to new observations. This paper examines the possibility of an anomaly in the distribution of galaxies rotating in opposite directions in the JWST deep fields as observed from Earth. The observation is compared to analyses with other space-based and Earth-based telescopes that image the same field, as well as other parts of the entire sky. 



\section{Data}
\label{data}

The data used in this analysis is taken from the GOODS-S field of JWST Advanced Deep Extragalactic Survey (JADES). JADES \citep{bunker2024jades,eisenstein2023overview} is the largest deep field imaging program planned for the early operation of JWST, focusing on the well-studied GOODS-S and GOODS-N fields. The image data are acquired primarily through the Near-Infrared Camera (NIRCam). 

The image data used for the analysis was based on the JWST 4.4um, 2.0um, and 0.9um bands visualized through the RGB channels, providing and informative form of visualization that allows for effective analysis. Parts of the JADES GOODS-S field that did not have these three channels were not used in the analysis. The RA of the objects used in this study ranged from $53.01885^o$ to $53.2184^o$, and the declination ranged between $-27.9145^o$ to $-27.7292^o$.

The galaxies were annotated by their direction of rotation as done in \citep{Shamir_2024}. The analysis was automatic, and followed a defined model that allows to define the direction of rotation of a galaxy in an objective and consistent manner. As also briefly mentioned in Section~\ref{introduction}, manual annotation of the direction of rotation of a galaxy can be subjective, and different people might have different opinions when they need to determine the direction of rotation of a galaxy. It has also been shown that such annotation can be driven by consistent biases, so even a group of people annotating the same galaxy cannot provide consistent annotations in all cases \citep{land2008galaxy}. For that reason, while manual annotation should be used to verify the consistency of the annotation, it cannot be considered a sound scientific methodology that such annotation can rely on.

Deep convolutional neural networks (CNNs) have become the most common solution to tasks related to image classification tasks. Their popularity is driven by excellent performance, as well as the availability of easy-to-use libraries that make the analysis accessible also to researchers who do not necessarily have strong computing skills. The primary downside of CNNs is that they are driven by highly complex data-driven rules that are very difficult to understand. Therefore, they are subjected to biases that can be highly difficult to identify \citep{DHAR202192,ball2023,electronics13101839}. Such biases can be driven by the manual selection of training samples, as two neural networks trained with two different training sets will also perform differently. In the case of astronomical images, even the distribution of the training galaxy images in the sky can lead to different results \citep{DHAR2022100545}. Therefore, using CNNs for the annotation of the galaxies cannot be considered a sound solution when the analysis needs to be clear, and certain conditions such as the symmetry of the algorithm need to be guaranteed.  

\begin{figure}
\centering
\includegraphics[scale=0.6]{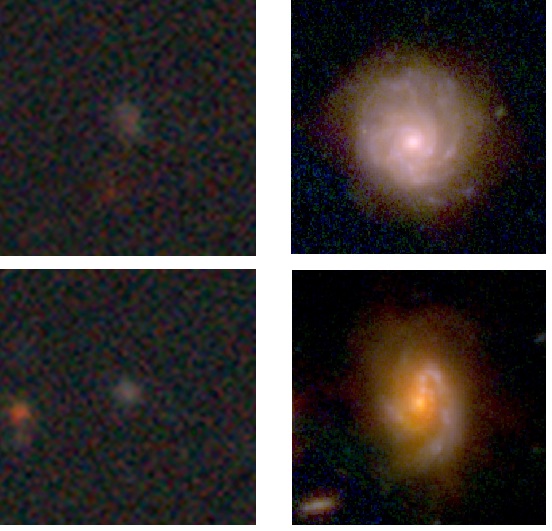}
\caption{Example of the same galaxies imaged by DES (left) and by JWST. JWST allows to analyze galaxies that DES or other Earth-based telescopes cannot image with sufficient details to identify their direction of rotation.}
\label{jwst_des}
\end{figure}

Clearly, the annotation of the galaxies is not complete, as some of the galaxies are not assigned with a direction of rotation, and are therefore excluded from the analysis. Some of the galaxies may be elliptical, and other galaxies may not have clear visual details that are sufficient to identify their direction of rotation. Figure~\ref{jwst_des} shows examples of galaxies imaged by both DES and JWST. While the DES images do not provide sufficient visual details, the JWST images allow to identify the direction of rotation of the galaxies. Therefore, JWST can provide an analysis of the direction of rotation of galaxies that cannot be imaged by DES or by any other existing Earth-based telescope.

The annotations were done by using the Ganalyzer algorithm \citep{shamir2011ganalyzer,ganalyzer_ascl} and also used in \citep{shamir2011ganalyzer,shamir2013color,shamir2016asymmetry,shamir2017photometric,shamir2017large,shamir2017colour,shamir2020large,shamir2020pasa,shamir2021particles,shamir2021large,shamir2022new,shamir2022large,shamir2022analysis, Shamir_2024}. The algorithm is based on a first step of separating foreground objects from the background. 
After each object is separated from the image, it is transformed into its radial intensity plot transformation.

The radial intensity plot captures the object intensity variations at different distances from its center. It is a 35$\times$360 matrix, where the intensity of the pixel $(x,y)$ is the median intensity of the 5$\times$5 pixels centered at $(O_x+\sin(\theta) \cdot r,O_y-\cos(\theta)\cdot r)$ in the original image, where {\it r} is the radial distance from the centre $(O_x,O_y)$, and $\theta$ is the polar angle \citep{shamir2011ganalyzer}.

\begin{figure*}
\centering
\includegraphics[scale=0.75]{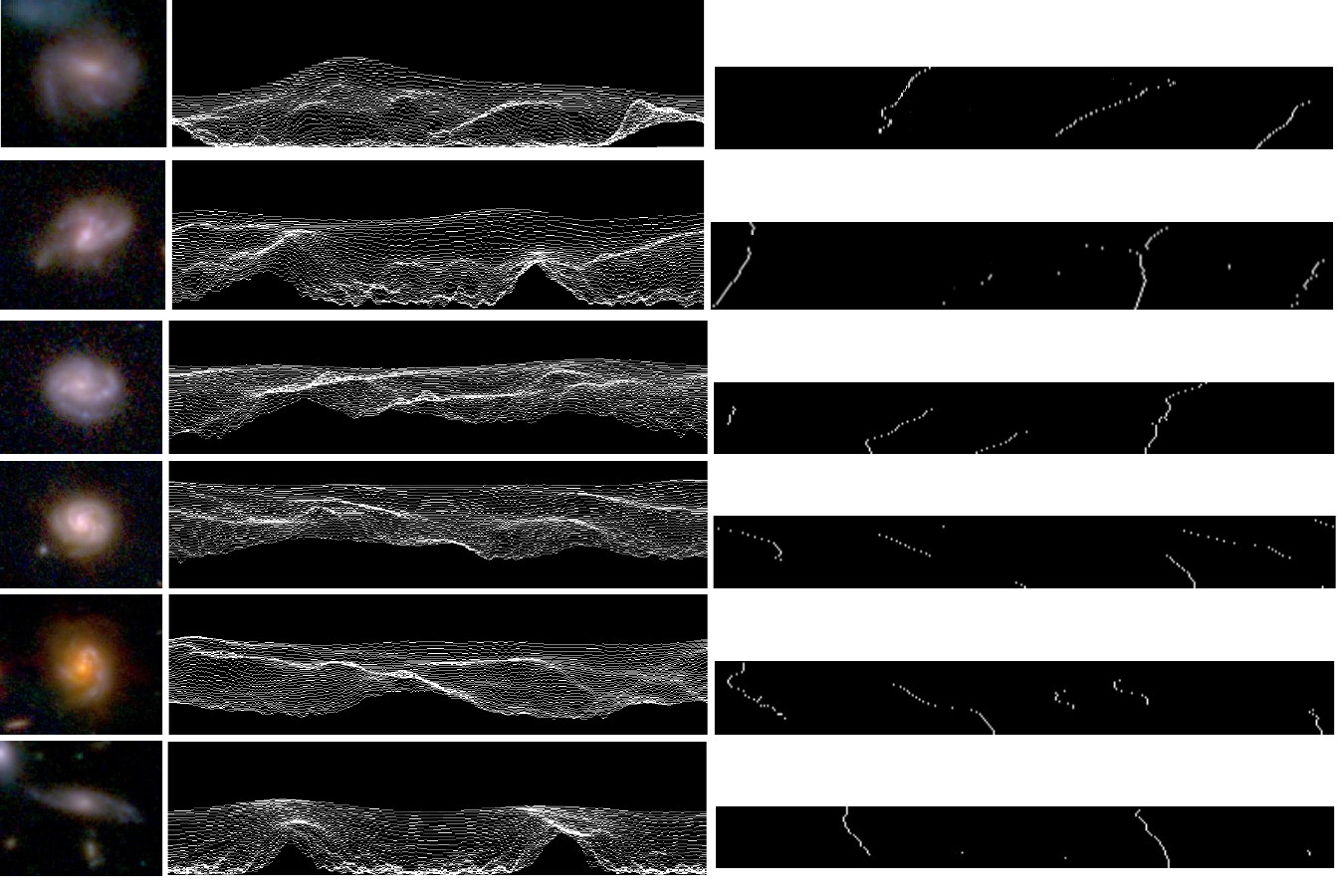}
\caption{Example of galaxies imaged by JWST and the peaks of the radial intensity plot transformations of each image. The lines formed by the peaks allow to identify the direction of the curve of the arms, and consequently the spin direction of the galaxy.}
\label{radials}
\end{figure*}

Because the arms of a galaxy are brighter than the non-arm part of the galaxy at the same distance from the galaxy centre, the arm pixels can be identified by a peak detection algorithm \citep{morhavc2000identification} applied to each line in the radial intensity plot. Applying a linear regression to the peaks provides the slope of the line formed by them, and the sign of the slope determines the direction towards which the arm of the galaxy is curved. To avoid elliptical galaxies or galaxies that do not have a clear direction of rotation, galaxies that have less than 30 peaks are considered galaxies that do not have an identifiable direction of rotation. Also, the slope of the linear regression needs to be at least 0.35 \citep{shamir2011ganalyzer}, otherwise the galaxy is not assigned with a direction of rotation, and therefore is not used in the analysis. Figure~\ref{radials} shows examples of galaxies as imaged by JWST as used in this study, and their radial intensity plots that allow to identify their direction of rotation. The process is described with empirical analysis and experimental results in \citep{shamir2011ganalyzer,shamir2013color,shamir2016asymmetry,shamir2017photometric,shamir2017large,shamir2017colour,shamir2020large,shamir2020pasa,shamir2021particles,shamir2021large,shamir2022new,shamir2022large,shamir2022analysis}.   Namely, it has been used to analyze initial JWST deep field images taken inside the footprint of GOODS-S \citep{Shamir_2024}.

\begin{figure}
\centering
\includegraphics[scale=0.7]{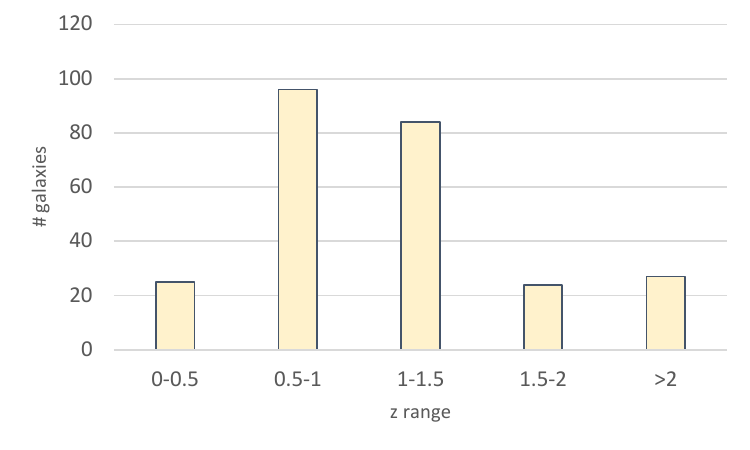}
\caption{The redshift distribution of the JWST galaxies used in the study.}
\label{redshift}
\end{figure}

The process led to 263 galaxies with identified direction of rotation. 
Figure~\ref{redshift} shows the distribution of the redshift of the galaxies.

While the algorithm is defined and symmetric, to ensure the symmetric nature of the analysis the entire field was mirrored, and the algorithm was applied to the mirrored image. Results were exactly inverse, which can be expected since the algorithm is symmetric, and was tested in a similar manner in previous experiments \citep{shamir2011ganalyzer,shamir2013color,shamir2016asymmetry,shamir2017photometric,shamir2017large,shamir2017colour,shamir2020large,shamir2020pasa,shamir2021particles,shamir2021large,shamir2022new,shamir2022large,shamir2022analysis, Shamir_2024}.

The annotation algorithm determines the directions of rotation of the galaxies by the curves of the arms. The arms of spiral galaxies have been shown to be a very reliable probe for determining the direction of rotation of the stellar mass as it rotates around the galaxy centre. For instance, \cite{de1958tilt} used the galaxy spectra and dust silhouette to study the link between the direction of rotation and shape of the galaxy arms, and found that in all cases the spiral arms were trailing. A more recent study \citep{iye2019spin} also found that all galaxies that were examined have trailing arms, and therefore the shape of the arm is a strong indication of the direction of rotation of the stellar mass. In some very rare cases galaxies can also have leading arms. A known example of a galaxy with leading arms is NGC 4622 \citep{freeman1991simulating,byrd2019ngc}, but these cases are extremely rare.

\section{Results}
\label{results}

The application of the image processing to the JWST GOODS-S image data as described in Section~\ref{data} provided annotations for 263 galaxies that their direction of rotation was identified. Of these galaxies, 105 rotate counterclockwise, while 158 rotate clockwise. Assuming that the probability of a galaxy to rotate in a certain direction is completely random, the one-tailed binomial distribution probability to have such asymmetry or stronger by chance is $\sim$0.0007, which is $\sim$3.39$\sigma$.

\begin{figure*}
\centering
\includegraphics[scale=0.62]{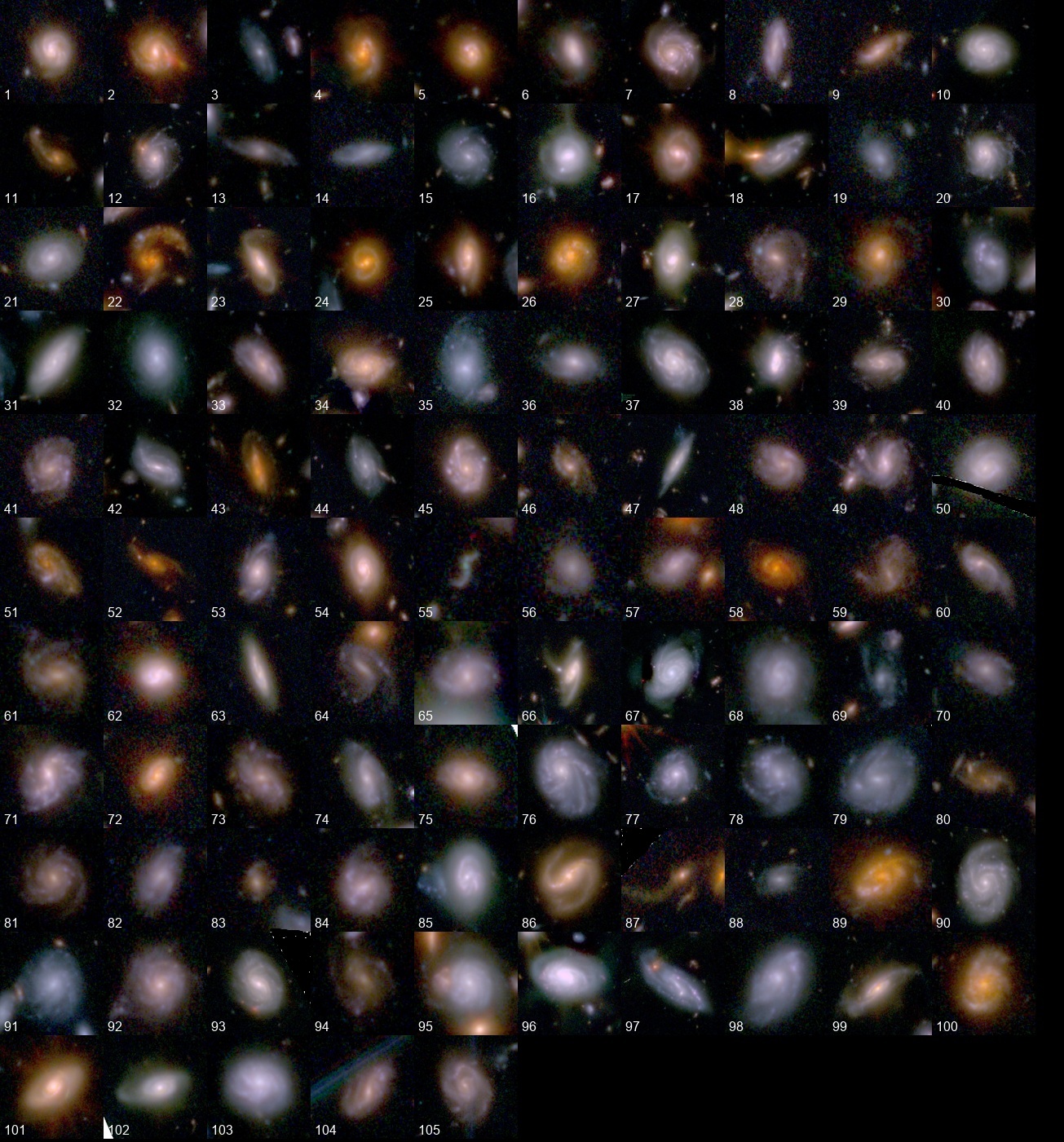}
\caption{The galaxies in the JADES GOODS-S field that were identified as rotating in the same direction relative to the Milky Way (counterclockwise). The $(\alpha,\delta)$ coordinates of each galaxy are specified in Table~\ref{ccwtab}.}
\label{ccw}
\end{figure*}

\begin{figure*}
\centering
\includegraphics[scale=0.53]{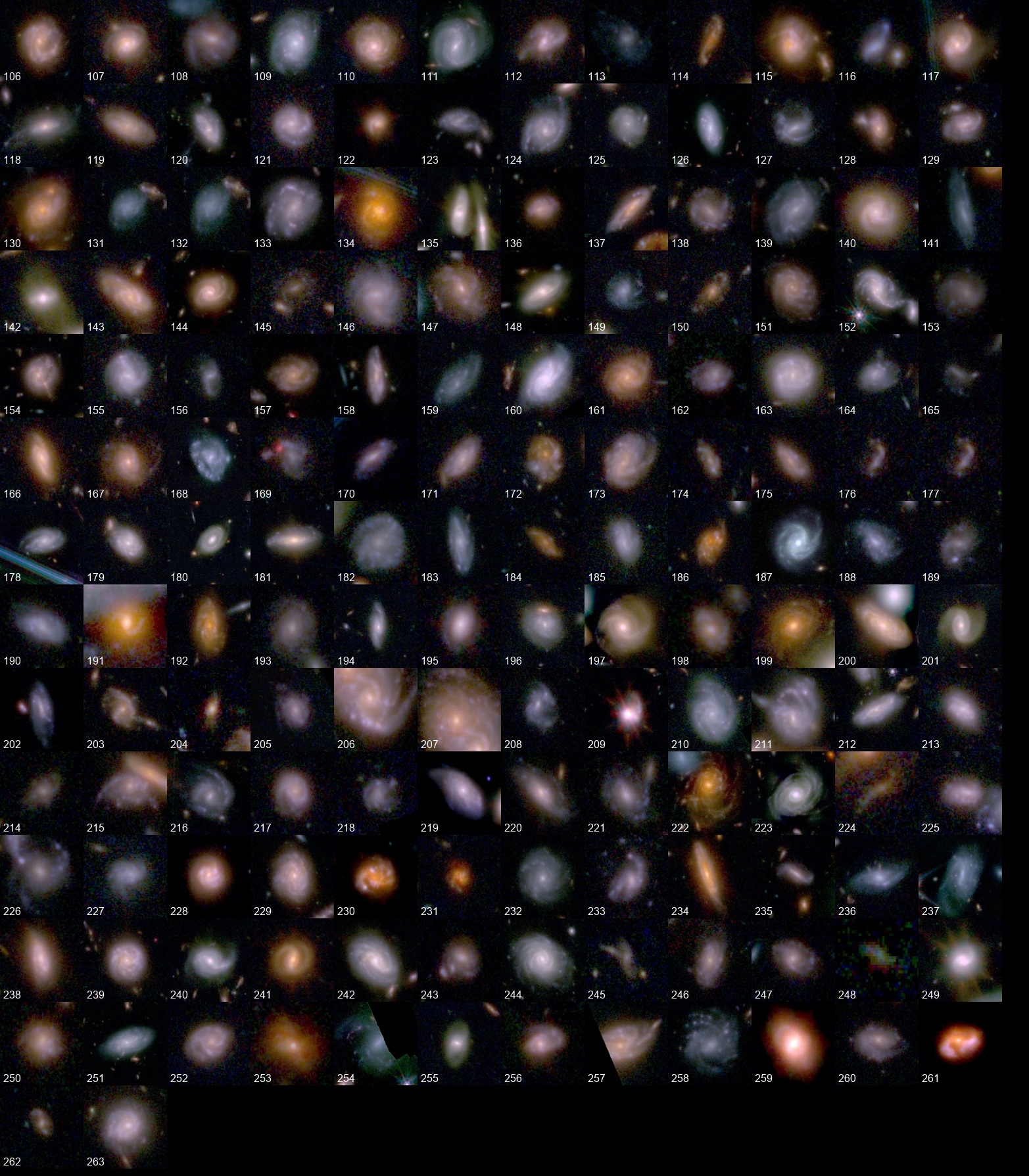}
\caption{The galaxies in the JADES GOODS-S field that were identified as rotating in the opposite direction relative to the Milky Way. The coordinates of each galaxy are specified in Table~\ref{cwtab}.}
\label{cw}
\end{figure*}

\begin{figure*}
\centering
\includegraphics[scale=0.06]{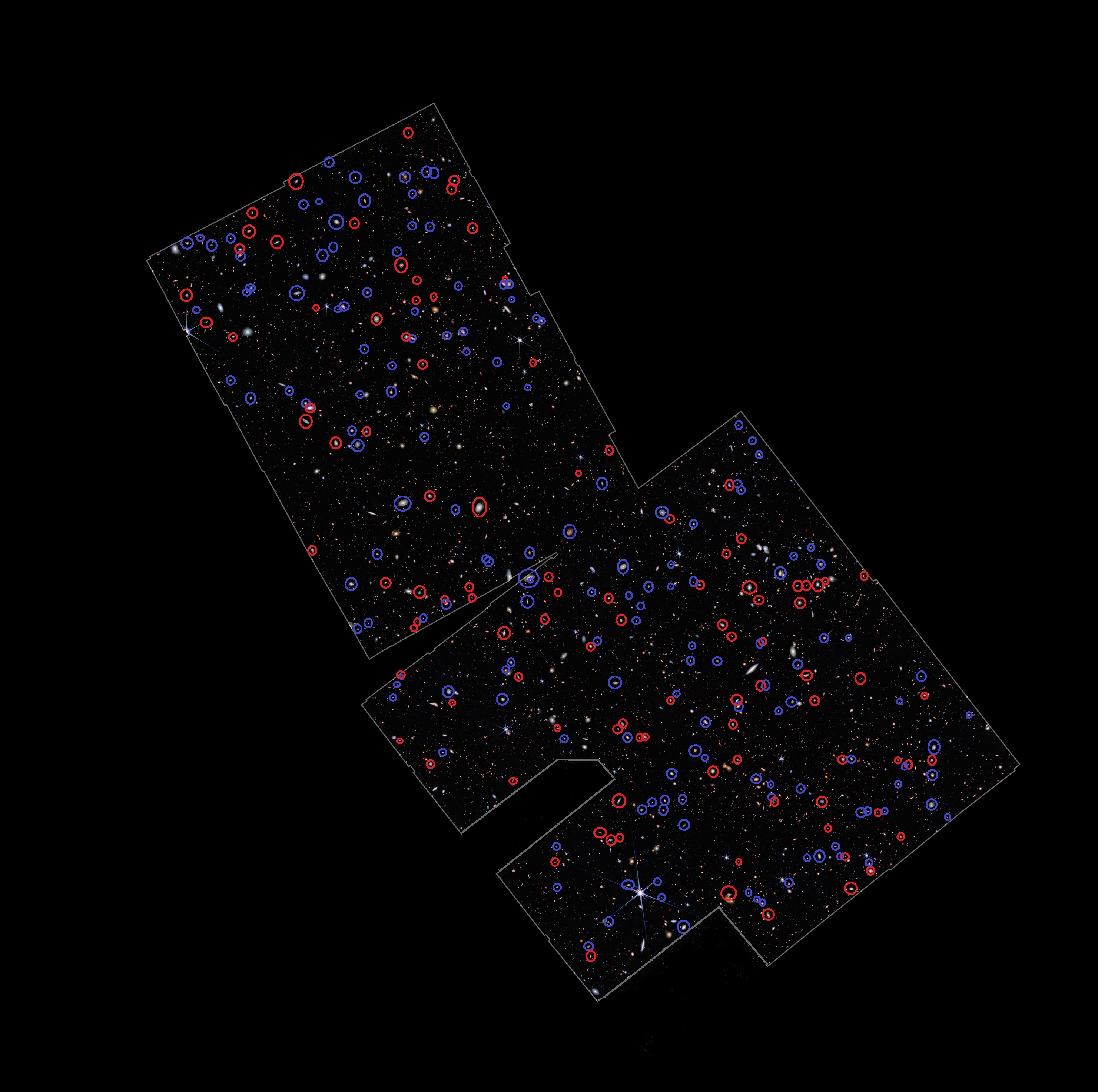}
\caption{Spiral galaxies imaged by JWST in the GOODS-S field of JADES that rotate in the same direction relative to the Milky Way (red), and in the opposite direction relative to the Milky Way (blue). The figure shows 158 galaxies that rotate in the opposite direction relative to the Milky Way, and just 105 that rotate in the same direction relative to the Milky Way. The analyzed field covers the JWST GOODS-S JADES field imaged with the 4.4um, 2.0um, and 0.9um bands.}
\label{annotated}
\end{figure*}

Figure~\ref{ccw} and~\ref{cw} show the galaxies in the field that were identified as rotating in the same direction relative to the Milky Way (counterclockwise) and the galaxies that rotate in the opposite direction relative to the Milky Way (clockwise), respectively. Tables~\ref{ccwtab} and~\ref{cwtab} provide the coordinates of each of the 263 galaxies. Figure~\ref{annotated} shows the location of the galaxies inside the JADES GOODS-S field. While the unaided human eye might not be a fully sound tool to annotate the galaxies, visual inspection shows consistency between the annotation of the algorithm and the human eye, and no galaxy seems to be identified incorrectly.

\begin{table*}
\centering
\small
\begin{tabular}{lcccccccccc}
\hline
\# & RA & Dec & & \# & RA & Dec & & \# & RA & Dec \\
\hline
1 & 53.0287575 & -27.8703469 & & 2 & 53.0305730 & -27.8557305 & & 3 & 53.0346279 & -27.8713092 \\
4 & 53.0367451 & -27.8875751 & & 5 & 53.0373338 & -27.8703302 & & 6 & 53.0425003 & -27.8821344 \\
7 & 53.0442441 & -27.8953532 & & 8 & 53.0460538 & -27.8289356 & & 9 & 53.0467626 & -27.8520855 \\
10 & 53.0495568 & -27.8992607 & & 11 & 53.0508248 & -27.8919429 & & 12 & 53.0514024 & -27.8701501 \\
13 & 53.0552519 & -27.8856482 & & 14 & 53.0558147 & -27.8300434 & & 15 & 53.0567343 & -27.8796932 \\
16 & 53.0578412 & -27.8307351 & & 17 & 53.0585653 & -27.8568762 & & 18 & 53.0605633 & -27.8512857 \\
19 & 53.0606090 & -27.8310224 & & 20 & 53.0625037 & -27.8349052 & & 21 & 53.0629972 & -27.8310975 \\
22 & 53.0689231 & -27.8796888 & & 23 & 53.0704161 & -27.9052013 & & 24 & 53.0717760 & -27.8437142 \\
25 & 53.0720162 & -27.8537413 & & 26 & 53.0727344 & -27.8343114 & & 27 & 53.0752012 & -27.8314599 \\
28 & 53.0772016 & -27.8205387 & & 29 & 53.0778957 & -27.8932042 & & 30 & 53.0781977 & -27.8701926 \\
31 & 53.0782543 & -27.8569828 & & 32 & 53.0794105 & -27.8623380 & & 33 & 53.0796836 & -27.8424121 \\
34 & 53.0803383 & -27.9005456 & & 35 & 53.0803688 & -27.8083994 & & 36 & 53.0810023 & -27.8238445 \\
37 & 53.0819773 & -27.8399439 & & 38 & 53.0844410 & -27.8728831 & & 39 & 53.0878515 & -27.8308227 \\
40 & 53.0951927 & -27.8568509 & & 41 & 53.0956144 & -27.8159647 & & 42 & 53.1018111 & -27.8650500 \\
43 & 53.1031616 & -27.8652455 & & 44 & 53.1074863 & -27.8623169 & & 45 & 53.1077205 & -27.8388817 \\
46 & 53.1081399 & -27.8877607 & & 47 & 53.1083292 & -27.8795013 & & 48 & 53.1085802 & -27.8633293 \\
49 & 53.1103699 & -27.8883061 & & 50 & 53.1107677 & -27.8339117 & & 51 & 53.1108220 & -27.8006488 \\
52 & 53.1131144 & -27.8866044 & & 53 & 53.1155528 & -27.9144036 & & 54 & 53.1155789 & -27.8447307 \\
55 & 53.1187014 & -27.8057706 & & 56 & 53.1237878 & -27.8326410 & & 57 & 53.1239903 & -27.8631171 \\
58 & 53.1245910 & -27.8932989 & & 59 & 53.1262410 & -27.8292075 & & 60 & 53.1272777 & -27.8386627 \\
61 & 53.1301701 & -27.7809458 & & 62 & 53.1326939 & -27.8329912 & & 63 & 53.1338997 & -27.8516127 \\
64 & 53.1352184 & -27.8750690 & & 65 & 53.1373614 & -27.7622325 & & 66 & 53.1374871 & -27.8418249 \\
67 & 53.1437771 & -27.8134799 & & 68 & 53.1454738 & -27.7506179 & & 69 & 53.1454750 & -27.8336683 \\
70 & 53.1462674 & -27.8314869 & & 71 & 53.1499913 & -27.7399957 & & 72 & 53.1507510 & -27.8574200 \\
73 & 53.1508711 & -27.7419219 & & 74 & 53.1523853 & -27.8343908 & & 75 & 53.1554348 & -27.7661381 \\
76 & 53.1564312 & -27.8108991 & & 77 & 53.1564439 & -27.8710278 & & 78 & 53.1581806 & -27.7811417 \\
79 & 53.1589977 & -27.8326493 & & 80 & 53.1595345 & -27.8392275 & & 81 & 53.1598219 & -27.7623367 \\
82 & 53.1600056 & -27.7669025 & & 83 & 53.1603471 & -27.8406146 & & 84 & 53.1618000 & -27.7292176 \\
85 & 53.1624406 & -27.7751180 & & 86 & 53.1635603 & -27.7589436 & & 87 & 53.1637381 & -27.8514252 \\
88 & 53.1643375 & -27.8658913 & & 89 & 53.1675906 & -27.8304041 & & 90 & 53.1698898 & -27.7710453 \\
91 & 53.1726089 & -27.7964452 & & 92 & 53.1754466 & -27.7496166 & & 93 & 53.1801809 & -27.7989112 \\
94 & 53.1852851 & -27.7685314 & & 95 & 53.1862976 & -27.8230718 & & 96 & 53.1868309 & -27.7910178 \\
97 & 53.1879238 & -27.7940122 & & 98 & 53.1902891 & -27.7401635 & & 99 & 53.1951860 & -27.7538285 \\
100 & 53.2015172 & -27.7472044 & & 101 & 53.2023884 & -27.7513782 & & 102 & 53.2046668 & -27.7553699 \\
103 & 53.2064105 & -27.7750928 & & 104 & 53.2131711 & -27.7718123 & & 105 & 53.2181259 & -27.7658299 \\
\hline
\end{tabular}
\caption{The RA and Declination of the galaxies that rotate counterclockwise shown in Figure~\ref{ccw}.}
\label{ccwtab}
\end{table*}

\begin{table*}
\centering
\small
\begin{tabular}{lcccccccccc}
\hline
\# & RA & Dec & & \# & RA & Dec & & \# & RA & Dec \\
\hline
106 & 3.1279542 & -27.7715134 & & 107 & 53.0193353 & -27.8602004 & & 108 & 53.0248564 & -27.8832216 \\
109 & 53.0281374 & -27.8675260 & & 110 & 53.0286673 & -27.8737467 & & 111 & 53.0289134 & -27.8803077 \\
112 & 53.0315505 & -27.8515717 & & 113 & 53.0357380 & -27.8716810 & & 114 & 53.0368943 & -27.8572009 \\
115 & 53.0373588 & -27.8758075 & & 116 & 53.0408007 & -27.8817782 & & 117 & 53.0446664 & -27.8936859 \\
118 & 53.0450450 & -27.8818705 & & 119 & 53.0465296 & -27.8821387 & & 120 & 53.0492605 & -27.8700264 \\
121 & 53.0499228 & -27.8428186 & & 122 & 53.0521408 & -27.8920364 & & 123 & 53.0533498 & -27.8897312 \\
124 & 53.0561208 & -27.8427935 & & 125 & 53.0569792 & -27.8262655 & & 126 & 53.0573808 & -27.8919577 \\
127 & 53.0595572 & -27.8224721 & & 128 & 53.0604167 & -27.8924328 & & 129 & 53.0620101 & -27.8768390 \\
130 & 53.0627890 & -27.8489887 & & 131 & 53.0638817 & -27.8243501 & & 132 & 53.0639208 & -27.8243688 \\
133 & 53.0642316 & -27.8572046 & & 134 & 53.0650035 & -27.8981819 & & 135 & 53.0673265 & -27.8282897 \\
136 & 53.0678622 & -27.8592175 & & 137 & 53.0693205 & -27.8788218 & & 138 & 53.0697243 & -27.8758420 \\
139 & 53.0710796 & -27.8537283 & & 140 & 53.0717995 & -27.9025473 & & 141 & 53.0722817 & -27.8443486 \\
142 & 53.0727420 & -27.8012799 & & 143 & 53.0731238 & -27.9018674 & & 144 & 53.0734710 & -27.8746265 \\
145 & 53.0745397 & -27.7985159 & & 146 & 53.0753324 & -27.9002020 & & 147 & 53.0772981 & -27.8095869 \\
148 & 53.0779581 & -27.8582796 & & 149 & 53.0780646 & -27.7948371 & & 150 & 53.0782010 & -27.8081526 \\
151 & 53.0832777 & -27.8480718 & & 152 & 53.0862719 & -27.8617956 & & 153 & 53.0864800 & -27.8698430 \\
154 & 53.0888870 & -27.8681537 & & 155 & 53.0893112 & -27.8172391 & & 156 & 53.0893392 & -27.8302602 \\
157 & 53.0897244 & -27.8446916 & & 158 & 53.0902375 & -27.8479062 & & 159 & 53.0917752 & -27.8850705 \\
160 & 53.0917947 & -27.9079949 & & 161 & 53.0921738 & -27.8791953 & & 162 & 53.0937000 & -27.8554013 \\
163 & 53.0942337 & -27.8755992 & & 164 & 53.0951142 & -27.8262703 & & 165 & 53.0951248 & -27.8313279 \\
166 & 53.0966713 & -27.8793750 & & 167 & 53.0970239 & -27.8816290 & & 168 & 53.0971496 & -27.8146692 \\
169 & 53.0973401 & -27.9013593 & & 170 & 53.0985649 & -27.8976944 & & 171 & 53.0998786 & -27.8798943 \\
172 & 53.1007743 & -27.8312572 & & 173 & 53.1025891 & -27.8815702 & & 174 & 53.1027136 & -27.8357093 \\
175 & 53.1039113 & -27.8390226 & & 176 & 53.1058087 & -27.8334029 & & 177 & 53.1058206 & -27.8334083 \\
178 & 53.1058432 & -27.8984435 & & 179 & 53.1060740 & -27.8652313 & & 180 & 53.1073507 & -27.8267313 \\
181 & 53.1091200 & -27.8530365 & & 182 & 53.1109910 & -27.9067594 & & 183 & 53.1125336 & -27.8080579 \\
184 & 53.1137720 & -27.8435787 & & 185 & 53.1152094 & -27.8325756 & & 186 & 53.1160406 & -27.9121885 \\
187 & 53.1207774 & -27.8189708 & & 188 & 53.1221556 & -27.8654683 & & 189 & 53.1242332 & -27.8897040 \\
190 & 53.1293315 & -27.7709544 & & 191 & 53.1308698 & -27.8299421 & & 192 & 53.1310120 & -27.8236125 \\
193 & 53.1315421 & -27.7864976 & & 194 & 53.1316813 & -27.8345866 & & 195 & 53.1356372 & -27.7666907 \\
196 & 53.1357641 & -27.8484238 & & 197 & 53.1362945 & -27.7632657 & & 198 & 53.1369272 & -27.7907591 \\
199 & 53.1370804 & -27.8501197 & & 200 & 53.1376604 & -27.7632512 & & 201 & 53.1378387 & -27.8566923 \\
202 & 53.1392775 & -27.7807792 & & 203 & 53.1413527 & -27.8257550 & & 204 & 53.1418679 & -27.8253231 \\
205 & 53.1470842 & -27.7785246 & & 206 & 53.1479323 & -27.7740950 & & 207 & 53.1481753 & -27.7738463 \\
208 & 53.1492580 & -27.7636845 & & 209 & 53.1499215 & -27.8140455 & & 210 & 53.1515470 & -27.8549017 \\
211 & 53.1519920 & -27.7747388 & & 212 & 53.1520962 & -27.8351648 & & 213 & 53.1531176 & -27.8686182 \\
214 & 53.1554066 & -27.7382866 & & 215 & 53.1572340 & -27.7379451 & & 216 & 53.1578600 & -27.7975775 \\
217 & 53.1580437 & -27.8384650 & & 218 & 53.1602537 & -27.7693798 & & 219 & 53.1607121 & -27.7753929 \\
220 & 53.1608299 & -27.7500577 & & 221 & 53.1608511 & -27.7428529 & & 222 & 53.1627502 & -27.7391179 \\
223 & 53.1631045 & -27.8123967 & & 224 & 53.1636135 & -27.8516310 & & 225 & 53.1646982 & -27.8533656 \\
226 & 53.1648402 & -27.7560441 & & 227 & 53.1657665 & -27.8562203 & & 228 & 53.1658978 & -27.7816064 \\
229 & 53.1661972 & -27.7875825 & & 230 & 53.1697386 & -27.8239961 & & 231 & 53.1719913 & -27.8395618 \\
232 & 53.1721962 & -27.7651716 & & 233 & 53.1729385 & -27.7779153 & & 234 & 53.1729692 & -27.7446003 \\
235 & 53.1740883 & -27.7881150 & & 236 & 53.1747265 & -27.8408071 & & 237 & 53.1747682 & -27.7992826 \\
238 & 53.1753216 & -27.7393471 & & 239 & 53.1762360 & -27.7962420 & & 240 & 53.1763802 & -27.8306977 \\
241 & 53.1765762 & -27.7855088 & & 242 & 53.1784236 & -27.7683139 & & 243 & 53.1796778 & -27.7688462 \\
244 & 53.1801045 & -27.7492538 & & 245 & 53.1808938 & -27.7549298 & & 246 & 53.1821319 & -27.7358393 \\
247 & 53.1835660 & -27.7568481 & & 248 & 53.1845989 & -27.7447883 & & 249 & 53.1879459 & -27.7900924 \\
250 & 53.1885059 & -27.7452762 & & 251 & 53.1901361 & -27.7652570 & & 252 & 53.1920889 & -27.7872543 \\
253 & 53.1922054 & -27.7410196 & & 254 & 53.2018256 & -27.7642261 & & 255 & 53.2018992 & -27.7888604 \\
256 & 53.2028329 & -27.7650603 & & 257 & 53.2047339 & -27.7568440 & & 258 & 53.2069528 & -27.7849266 \\
259 & 53.2070046 & -27.7529810 & & 260 & 53.2118685 & -27.7544650 & & 261 & 53.2146651 & -27.7526799 \\
262 & 53.2157378 & -27.7691409 & & 263 & 53.2180201 & -27.7540485 & &  & &  \\
\hline
\end{tabular}
\caption{The RA and Declination of the galaxies that rotate clockwise shown in Figure~\ref{cw}.}
\label{cwtab}
\end{table*}

Visual inspection of the galaxies that were identified shows no galaxy that was annotated incorrectly. But the annotation algorithm also avoids annotating galaxies that their direction of rotation cannot be identified. These may be elliptical galaxies, or galaxies that the visual details of their arms do not allow the identification of their direction of rotation. Because the algorithm is symmetric, galaxies that their direction of rotation could not be identified by the algorithm are expected to be treated in the same manner regardless of their direction of rotation. Still, it is important to also inspect the population of galaxies that the algorithm could not identify their direction of rotation.

\begin{figure}
\centering
\includegraphics[scale=0.35]{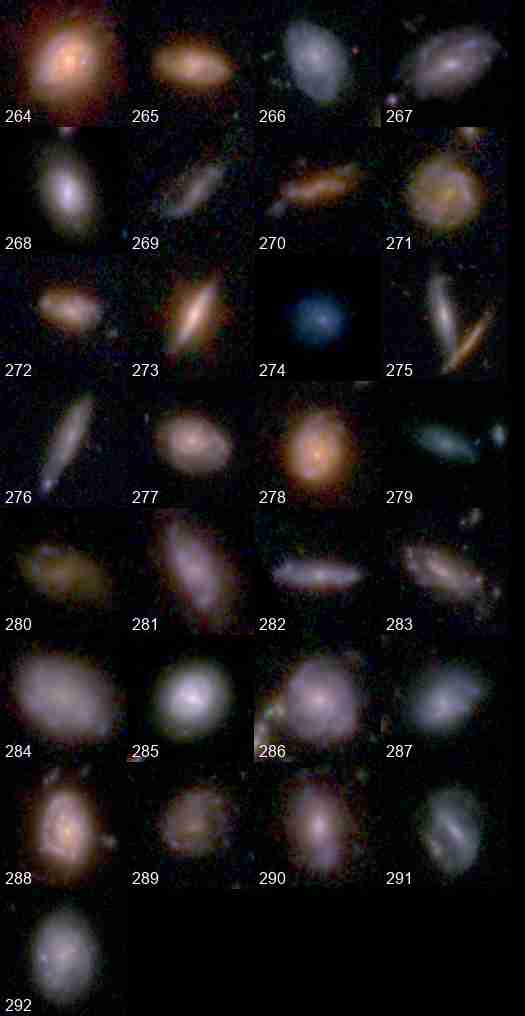}
\caption{The galaxies in the JADES GOODS-S field that were not identified by the annotation algorithm, but were identified through manual inspection as galaxies that could be rotating in the same direction relative to the Milky Way (counterclockwise). The direction of rotation in these galaxies is not entirely clear from the images.}
\label{bccw}
\end{figure}

\begin{figure}
\centering
\includegraphics[scale=0.35]{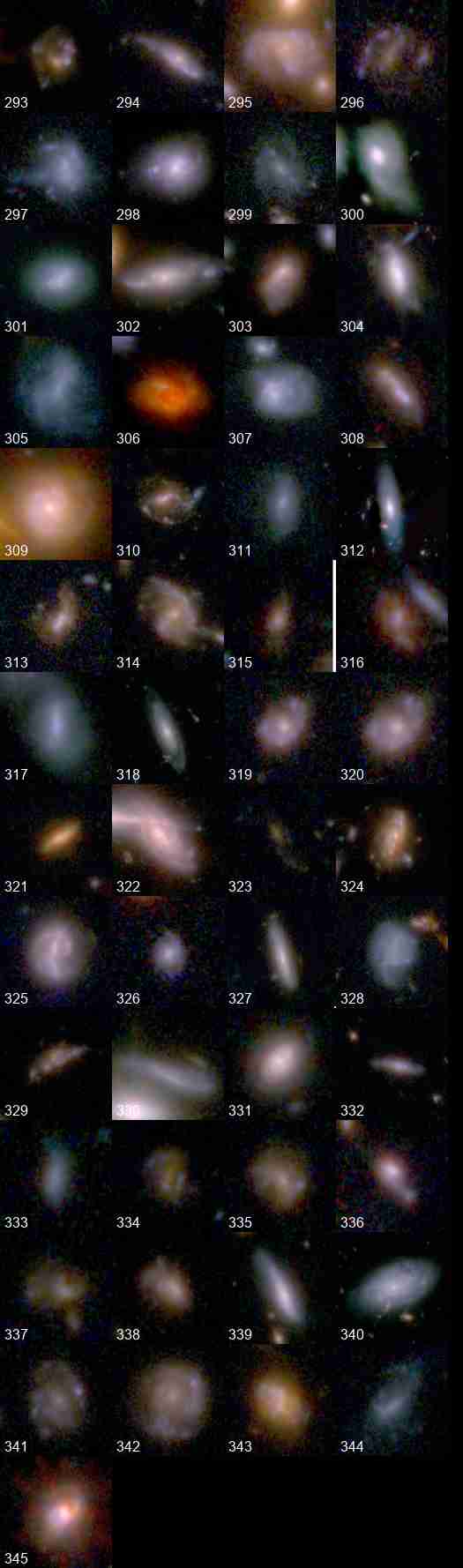}
\caption{The galaxies in the JADES GOODS-S field that were not identified by the annotation algorithm, but were identified manually as galaxies that could be rotating in the opposite direction relative to the Milky Way.}
\label{bcw}
\end{figure}

As was done in \citep{Shamir_2024}, the field was inspected manually to identify galaxies that their direction of rotation was not identified by the algorithm. Figures~\ref{bccw} and~\ref{bcw} show galaxies that perhaps could be considered as rotating counterclockwise and clockwise, respectively. The rotation of direction is not entirely clear in these images, but these galaxies were the most clear galaxies among those galaxies that were not identified by the algorithm. As mentioned before, manual observation is not a sound scientific manner to identify the direction of rotation of the galaxies due to its subjective nature. These galaxies were not used in the analysis, but just to inspect the kind of galaxies that were not annotated by the algorithm. The manual inspection does not show a certain pattern of galaxies that their direction of rotation was not identified by the algorithm. The coordinates of the galaxies in both figures are specified in Tables~\ref{bccwtab} and~\ref{bcwtab}.

\begin{table*}
\centering
\small
\begin{tabular}{lcccccccccc}
\hline
\# & RA & Dec & & \# & RA & Dec & & \# & RA & Dec \\
\hline
264 & 53.0384867 & -27.8622052 & & 265 & 53.0502416 & -27.8402054 & & 266 & 53.0571043 & -27.8209306 \\
267 & 53.0734400 & -27.8408036 & & 268 & 53.0784400 & -27.8194916 & & 269 & 53.0938238 & -27.8941047 \\
270 & 53.0960714 & -27.8286830 & & 271 & 53.0968709 & -27.8855638 & & 272 & 53.0979105 & -27.9080514 \\
273 & 53.1119073 & -27.8053042 & & 274 & 53.1205041 & -27.8517213 & & 275 & 53.1315577 & -27.8963155 \\
276 & 53.1329688 & -27.8984036 & & 277 & 53.1393988 & -27.7674556 & & 278 & 53.1411431 & -27.7618807 \\
279 & 53.1435404 & -27.7557315 & & 280 & 53.1500039 & -27.7577261 & & 281 & 53.1523614 & -27.7780246 \\
282 & 53.1588635 & -27.7574229 & & 283 & 53.1607380 & -27.8358968 & & 284 & 53.1644578 & -27.7659101 \\
285 & 53.1673684 & -27.8405377 & & 286 & 53.1675621 & -27.7925632 & & 287 & 53.1807708 & -27.7568397 \\
288 & 53.1812865 & -27.7656881 & & 289 & 53.1821604 & -27.8058583 & & 290 & 53.1898988 & -27.7413466 \\
291 & 53.2070911 & -27.7641427 & & 292 & 53.2093471 & -27.7609204 & &  & &  \\
\hline
\end{tabular}
\caption{The RA and Declination of the galaxies that rotate counterclockwise shown in Figure~\ref{bccw}.}
\label{bccwtab}
\end{table*}

\begin{table*}
\centering
\small
\begin{tabular}{lcccccccccc}
\hline
\# & RA & Dec & & \# & RA & Dec & & \# & RA & Dec \\
\hline
293 & 53.0307616 & -27.8706095 & & 294 & 53.0331686 & -27.8480343 & & 295 & 53.0477066 & -27.8353006 \\
296 & 53.0552493 & -27.8233445 & & 297 & 53.0608612 & -27.8204453 & & 298 & 53.0689281 & -27.8263187 \\
299 & 53.0697976 & -27.8391838 & & 300 & 53.0711214 & -27.8227704 & & 301 & 53.0741950 & -27.8239991 \\
302 & 53.0882270 & -27.8506168 & & 303 & 53.0923399 & -27.8479983 & & 304 & 53.0965038 & -27.9073124 \\
305 & 53.1072662 & -27.8058498 & & 306 & 53.1074781 & -27.8041212 & & 307 & 53.1082963 & -27.8930911 \\
308 & 53.1184803 & -27.8053588 & & 309 & 53.1198685 & -27.7987721 & & 310 & 53.1228467 & -27.8483623 \\
311 & 53.1284391 & -27.8504606 & & 312 & 53.1360779 & -27.8292005 & & 313 & 53.1403351 & -27.8635211 \\
314 & 53.1413943 & -27.8257229 & & 315 & 53.1419006 & -27.8253096 & & 316 & 53.1436493 & -27.7582272 \\
317 & 53.1452409 & -27.7511319 & & 318 & 53.1508174 & -27.7601240 & & 319 & 53.1508706 & -27.8612319 \\
320 & 53.1508807 & -27.8611853 & & 321 & 53.1547878 & -27.7739138 & & 322 & 53.1548761 & -27.8578577 \\
323 & 53.1554620 & -27.8386084 & & 324 & 53.1557427 & -27.7794153 & & 325 & 53.1588282 & -27.7705852 \\
326 & 53.1600134 & -27.8637361 & & 327 & 53.1602644 & -27.8255784 & & 328 & 53.1608377 & -27.8653308 \\
329 & 53.1639085 & -27.7653675 & & 330 & 53.1692446 & -27.7917806 & & 331 & 53.1707031 & -27.7512668 \\
332 & 53.1713223 & -27.7930262 & & 333 & 53.1729254 & -27.7387797 & & 334 & 53.1763369 & -27.8251897 \\
335 & 53.1788946 & -27.7547901 & & 336 & 53.1814716 & -27.8318782 & & 337 & 53.1818135 & -27.8306672 \\
338 & 53.1825614 & -27.8244443 & & 339 & 53.1830999 & -27.7510514 & & 340 & 53.1901582 & -27.7652015 \\
341 & 53.1907319 & -27.7570106 & & 342 & 53.1926134 & -27.7581417 & & 343 & 53.2021859 & -27.7550068 \\
344 & 53.2023063 & -27.7904402 & & 345 & 53.2180747 & -27.7616774 & &  & &  \\
\hline
\end{tabular}
\caption{The RA and Declination of the galaxies that rotate clockwise shown in Figure~\ref{bcw}.}
\label{bcwtab}
\end{table*}

\begin{figure*}
\centering
\includegraphics[scale=0.5]{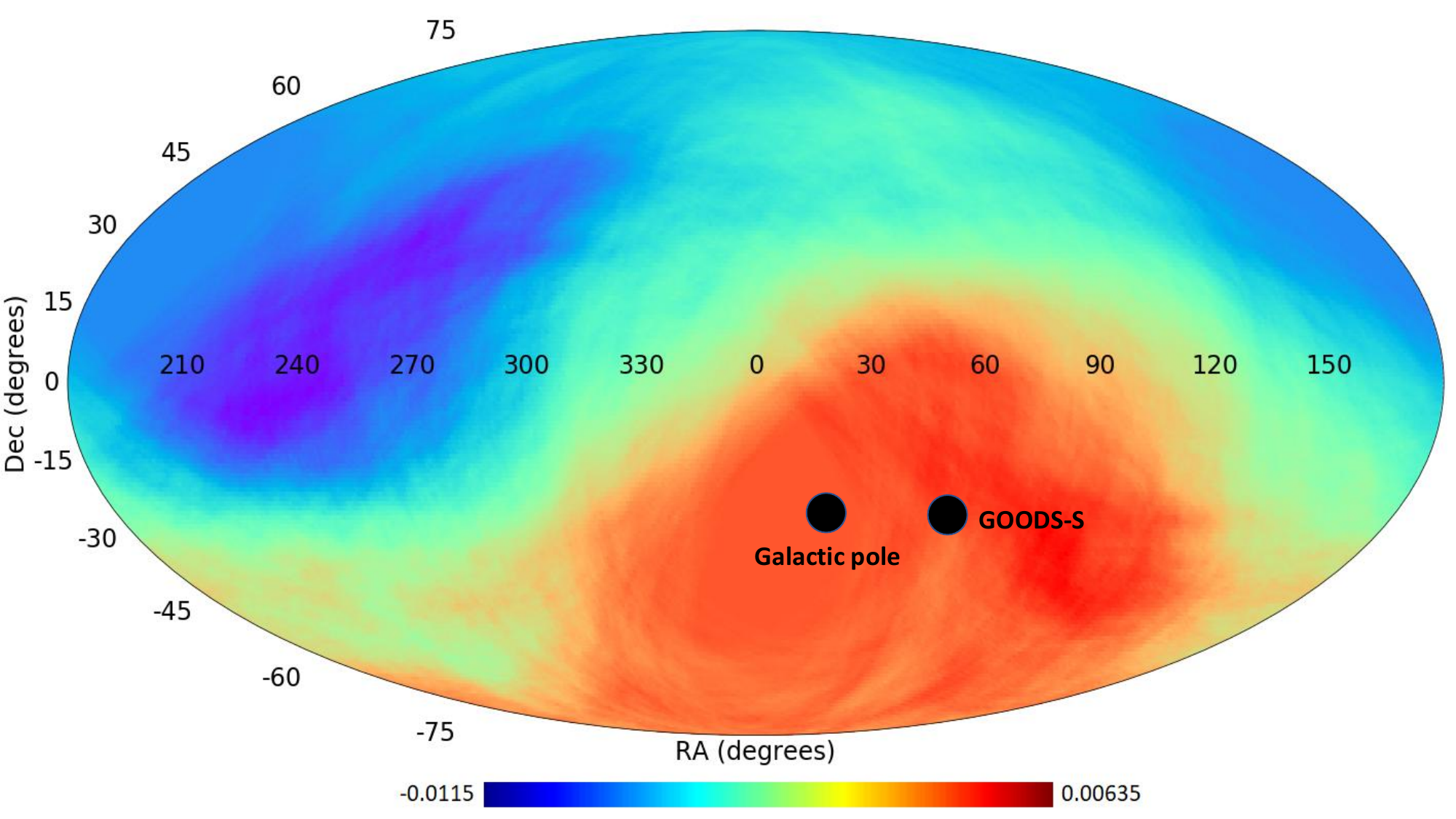}
\caption{The differences in the number of galaxies with opposite directions of rotations in different parts of the sky as determined by using $1.3\cdot10^6$ galaxies imaged by the DESI Legacy Survey \citep{shamir2022analysis}. The location of the GOODS-S field is at a part of the sky with a higher number of galaxies rotating clockwise.}
\label{jwst_location}
\end{figure*}

JWST provides visual details of galaxies in the deep Universe, and far deeper than any other telescope. But the unequal number of galaxies that rotate in opposite directions around the Galactic poles was noticed also with Earth-based telescopes, although the differences were smaller than the difference observed in the deeper Universe using JWST. For instance, Figure~\ref{jwst_location} shows the difference between the number of galaxies with opposite directions of rotation in different parts of the sky, as determined by using a large dataset of $1.3\cdot10^6$ galaxies annotated by their direction of rotation \citep{shamir2022analysis}. The galaxy images were collected by the DESI Legacy Survey, and the analysis was done before JWST was launched. The difference between galaxies that rotate in opposite directions in the different parts of the sky are quantified by $\frac{cw-ccw}{cw+ccw}$ in the hemisphere centered at each integer $(\alpha,\delta)$ combination, and displayed by the colour such that red parts of the sky indicate a higher number of galaxies rotating clockwise, and blue parts of the sky reflect a higher number of galaxies rotating counterclockwise. The figure shows simple direct measurements of the differences in different parts of the sky, and not an attempt to fit the distribution to a certain pre-determined model. The image and the analysis through which it was generated are explained in full detail in \citep{shamir2022analysis}.

The figure shows a higher asymmetry in both ends of the Galactic poles, where in both end there is a higher number of galaxies that rotate in the opposite directions relative to the Milky Way galaxy. GOODS-S is located in relatively close proximity to the Southern Galactic pole, and therefore the difference can be expected based on previous observations made before JWST was launched. Previous observations using Earth-based telescopes also showed that the magnitude of the asymmetry increases as the redshift gets higher \citep{shamir2020patterns}. If that trend continues into the higher redshift ranges, it can also explain the higher asymmetry in the much higher redshift of the galaxies imaged by JWST. 

``Webb's First Deep Field'' was also tested in the same manner, providing no statistically significant asymmetry with 21 galaxies rotating counterclockwise, and 19 galaxies that rotate clockwise \citep{Shamir_2024}. That field is not close to neither ends of the Galactic pole, so asymmetry is not expected in that field based on the analysis done with DESI Legacy Survey \citep{shamir2022analysis} before JWST was launched.

\section{Summary of experiments suggesting that the distribution of the galaxy directions of rotation is random}
\label{other_reports}

Section~\ref{introduction} mentions multiple studies using several different space-based and Earth-based telescopes showing unequal distribution of the directions of rotation of galaxies \citep{macgillivray1985anisotropy,longo2011detection,shamir2012handedness,shamir2016asymmetry,shamir2019large,shamir2020patterns,shamir2020large,shamir2020pasa,shamir2021particles,shamir2021large,shamir2022new,shamir2022large,shamir2022asymmetry,shamir2022analysis}. Reports started as early as the 1980's \citep{macgillivray1985anisotropy}, and include Earth-based telescopes such as SDSS  \citep{shamir2019large,shamir2020patterns,shamir2021particles,shamir2022large}, Pan-STARRS \citep{shamir2020patterns}, the Dark Energy Survey \citep{shamir2022asymmetry}, and DESI Legacy Survey \citep{shamir2021large,shamir2022analysis}, as well as space-based telescopes such as HST \citep{shamir2020pasa} and JWST \citep{Shamir_2024}.

Section~\ref{introduction} also mentions previous reports suggesting fully random distribution of the directions of rotations of galaxies. Although none of these studies used high redshifts space-based telescopes, these observations can be considered to be in conflict with the observation of unequal distribution shown in Section~\ref{results}. Analyses of these experiments, including reproduction of the results, show that these experiments are in fact aligned with the non-random distribution as shown in Section~\ref{results}. Explanations of these experiments as well as code and data to reproduce them can be found in \citep{sym15091704}, and description of specific experiments can be found in \citep{psac058,mcadam2023reanalysis,shamir2022analysis}.

One of the early experiments \citep{iye1991catalog} tested the distribution by annotating the galaxies manually. Besides the limitations of possible systematic biases of manual annotations, manual annotation is also highly limited by the volume of data that can be processed. The resulting dataset only included 3,257 galaxies rotating clockwise and 3,268 galaxies rotating counterclockwise. As explained quantitatively in \citep{sym15091704,psac058,mcadam2023reanalysis,shamir2022analysis}, that dataset was far too small to show a statistically significant difference for galaxies at relatively low redshift. Experiments using Earth-based telescopes used far larger datasets \citep{sym15091704,psac058,mcadam2023reanalysis,shamir2022analysis}.

An experiment that received public attention was made by using anonymous volunteers to annotate the galaxies through the Internet \citep{land2008galaxy}. The use of a large number of human annotators provided a large number of annotations. The downside of the approach was that the annotations made by the volunteers had a substantial degree errors, making most annotated galaxies unusable due to the high level of disagreements between the annotators. More importantly, the human annotators had a bias towards galaxies that rotate counterclockwise, leading to an extreme bias of  $\sim$15\% in the resulting ``superclean'' dataset. That bias did not allow to determine whether the excessive number of galaxies that rotate counterclockwise is driven by the Universe or by the bias of the volunteers who annotated them.

After the bias was noticed, a new experiment was done by annotating the original galaxy images as well as the mirrored images. But because the bias was noticed only after a very high number of galaxies were already annotated, the dataset of the new experiment was relatively small, and included just  $\sim1.1\cdot10^4$ annotated galaxies. The results are displayed in Table 2 in \citep{land2008galaxy}. As also explained in \citep{sym15091704,shamir2022analysis,mcadam2023reanalysis,psac058,Shamir_2024}, the table shows a 1.5\% higher number of galaxies rotating counterclockwise in the first experiment, and 2.2\% in the second experiment. Due to the small number of galaxies the statistical significance was marginal, and becomes significant only when combining the two experiments \citep{sym15091704,psac058,mcadam2023reanalysis,Shamir_2024}. But the asymmetry agrees on both the direction and the magnitude with the asymmetry shown in \citep{shamir2020patterns}, which uses the same telescope and same footprint as the experiment of \cite{land2008galaxy}.

\cite{hayes2017nature} used automation to annotate a large number of galaxies from the same telescope and footprint used in \citep{land2008galaxy,shamir2020patterns}. The results are summarized in Table 2 in \cite{hayes2017nature}, showing consistent results of an excessive number of galaxies that rotate counterclockwise. The experiment that provided random distribution was an experiment done by selecting the spiral galaxies by applying machine learning. Interestingly, the selection of the galaxies by using machine learning was done such that features that correlate with the direction of rotation of the galaxies were identified and removed manually. As stated in \citep{hayes2017nature}, ``We choose our attributes to include some photometric attributes that were disjoint with those that Shamir (2016) found to be correlated with chirality, in addition to several SpArcFiRe outputs with all chirality information removed''. The discussion does not specify a reason for the decision to manually remove just these features. 

After removing these features manually, the analysis provided random distribution. But that can also be expected because removing just the features that can identify galaxy direction of rotation would naturally weaken any signal of unequal number of galaxies that rotate in opposite direction. That is explained in detail in \citep{mcadam2023reanalysis}, as well as in \citep{sym15091704}. Reproduction of the experiment of \cite{hayes2017nature} by using the exact same code and same data but without manually removing specific features showed a clear statistically significant asymmetry \citep{mcadam2023reanalysis}, in good agreement with the asymmetry observed with SDSS and other telescopes. The full reproduction of the experiment with code and data is described in \citep{mcadam2023reanalysis}.

Another experiment used image data taken from the Hyper Suprime-Cam (HCS), and annotated it automatically using a deep neural network  \citep{tadaki2020spin}. That data analysis provided 38,718 galaxies that rotating clockwise and 37,917 galaxies rotating counterclockwise. Using simple binomial distribution, the one-tailed mere chance probability of the difference is p$\simeq$0.0019. The higher number of galaxies that rotate clockwise agrees with the location of HCS footprint, which is closer to the Southern Galactic pole. As shown in \citep{shamir2022analysis} and in Figure~\ref{jwst_location}, a higher number of galaxies rotating clockwise is expected at around the Southern Galactic pole.

Since the deep neural network used for the annotation had a certain degree of error, and the error was higher than the asymmetry, the analysis was not considered statistically significant. But although the results cannot be considered a sound proof for the unequal distribution, they are in agreement with the other previous reports that show an unequal number of galaxies that rotate in opposite directions, and certainly do not conflict with them.

\cite{jia2023galaxy} used deep neural networks to study the distribution of galaxy directions of rotation using data collected by SDSS and DESI. Due to the error in the annotation of the neural network, the experiment was done with different accuracy thresholds to balance between the accuracy of the annotations and the size of the dataset. When increasing the accuracy threshold, the error in the annotation of the data decreases, but the size of the dataset gets smaller since fewer galaxies meet the threshold. 

The highest threshold used was 0.9, providing a dataset of 9,218 SDSS clockwise galaxies and 9,442 counterclockwise SDSS galaxies, as shown in Table 1 in \cite{jia2023galaxy}. The asymmetry of $\sim$2.4\% agrees with the asymmetry shown in \cite{shamir2020patterns}, which is based on the same sky survey and therefore similar footprint. The one-tailed probability for the observation to occur by mere chance is $\sim$0.05. That statistical significance is somewhat weaker than the statistical significance observed in \citep{shamir2020patterns}, which can be expected due to the much higher number of galaxies used in \citep{shamir2020patterns}. But although the p value is lower, it can still be considered statistically significant, and definitely not in conflict with the other observations as explained in \citep{Shamir_2024}.

The DESI Legacy Survey has a very large footprint that covers both hemispheres, and therefore it is difficult to predict the asymmetry as the asymmetry in one hemisphere is offset by the inverse asymmetry in the opposite hemispheres \citep{Shamir_2024}. The analysis of \cite{jia2023galaxy} found 11,649 clockwise galaxies and 11,919 counterclockwise galaxies. The probability of that asymmetry to occur by chance is 0.04. While the large footprint does not allow accurate analysis, the binomial distribution can be considered statistically significant, and does not conflict with the contention that the distribution of the directions of rotation of the galaxies is random.

Another experiment suggesting random distribution of galaxy directions of rotation claimed that the asymmetry shown in previous experiments is the result of ``duplicate objects'' in the dataset \citep{iye2020spin}. This experiment is explained in detail and full reproduction in \citep{psac058,sym15091704,shamir2022analysis}. In summary, the catalog used in \cite{iye2020spin} was used in \citep{shamir2017photometric} for photometric analysis only, and no claim for a dipole axis formed by the distribution of the directions of rotation of galaxies was made based on that dataset \citep{psac058,sym15091704,shamir2022analysis}.

More importantly, as shown in \citep{sym15091704}, the reproduction of the experiment using the same data and same analysis method used in \citep{iye2020spin} provides different results than the results shown in \citep{iye2020spin}, showing a statistically significant non-random distribution \cite{sym15091704}. Code and data that allows to easily reproduce the experiment are available at \footnote{\label{note1}\url{https://people.cs.ksu.edu/~lshamir/data/iye_et_al}}. The link also provides the description of the reproduction of the National Astronomical Observatory of Japan (NAOJ) to explain the difference between the reproduction and the results shown in the paper. In summary, the NAOJ explains that ``Because it is hard to verify the detail of simulations, we here calculate the analytic solution by Chandrasekhar (1943) which assumes uniform samples in the hemisphere.'' But the assumption of a uniform sample is not correct for SDSS, which covers just a part of the hemisphere, and the density of the galaxy population varies substantially within its footprint. That assumption is not mentioned in \citep{iye2020spin}, and in fact is not needed since the exact locations of all galaxies are well known.

Another study \citep{petal} that suggested that the distribution of the directions of rotation of galaxies is random was based on data used in previous studies \citep{longo2011detection,mcadam2023reanalysis}. Instead of using the standard binomial distribution statistics and simple $\chi^2$ statistics used in \citep{macgillivray1985anisotropy,longo2011detection,shamir2012handedness,shamir2016asymmetry,shamir2019large,shamir2020patterns,shamir2020large,shamir2020pasa,shamir2021particles,shamir2021large,shamir2022new,shamir2022large,shamir2022asymmetry,shamir2022analysis,mcadam2023reanalysis,sym15091704,Shamir_2024},  \cite{petal} propose a new complex ad-hoc statistical method. The downside of the new method is that it does not respond to asymmetry in the distribution of galaxy spin directions \citep{shamir2024reproducible,sym16101389}. As explained in \citep{shamir2024reproducible}, even in cases where synthetic asymmetry is added to the data to create a highly asymmetric dataset, the method still reports random distribution. 

For instance, \cite{petal} applied the new method to the dataset used in \citep{mcadam2023reanalysis}, annotated as ``GAN M'' in \citep{petal}. It is publicly available at \url{https://people.cs.ksu.edu/~lshamir/data/sparcfire/}. The dataset was annotated by the {\it SpArcFiRe} algorithm for the purpose of reproducing the results shown in \citep{hayes2017nature}, not necessarily to study the Universe due to the known bias in the annotation method. As noted in the appendix of \citep{hayes2017nature}, {\it SpArcFiRe} has a known bias, and therefore a dataset annotated by it is expected to show a difference between the number of galaxies rotating clockwise and the number of galaxies rotating counterclockwise. Indeed, the dataset of 139,852 galaxies annotated by {\it SpArcFiRe} is separated to 70,672 counterclockwise galaxies and 69,180 clockwise galaxies. Using standard binomial statistics, the two-tailed probability to have such asymmetry by chance is $\sim0.00006$. But as reported in the fourth row of Table 4 in \cite{petal}, the new statistical method provided a non-significant p-value of 0.25. The fact that applying the new method to an extremely asymmetric synthetic data shows a null-hypothesis universe with no statistically significant asymmetry, indicating that the new method does not guarantee to identify asymmetry \citep{petal}.

\cite{petal} also argued that the reproduction of previous results of \citep{longo2011detection} and \citep{mcadam2023reanalysis} provided different results than the results stated in these papers, as stated in Section 4.3 in \citep{petal}. That claim has also shown to be incorrect, with code and step-by-step instructions to easily reproduce the results of both papers. The full open code and step-by-step instructions can be found at \url{https://people.cs.ksu.edu/~lshamir/data/patel_desmond/}.

\section{Conclusion}
\label{conclusion}

JWST has provided unprecedented imaging power that allows to observe high visual details of galaxies in the deep Universe. Despite being relatively new, observations made in JWST deep fields have already challenged some of the foundational assumptions regarding the Universe. Here, JWST shows that the number of galaxies that rotate in the opposite direction relative to the Milky Way as observed from Earth is higher than the number of galaxies that rotate in the same direction relative to the Milky Way. The observation was made also with initial JWST data \citep{Shamir_2024}, and was also noticed in the Ultra Deep Field imaged by HST \citep{shamir2021aas}, but JADES allows to observe the asymmetry in the early Universe using a much higher number of galaxies. The analysis is done by a defined quantitative criteria, but the asymmetry is high and can be inspected also by the unaided human eye.

As discussed in Section~\ref{introduction}, the asymmetry between galaxies that rotate in opposite directions was noticed in numerous studies starting the 1980s, and more recent experiments include analysis of very large datasets collected by autonomous digital sky surveys. The magnitude of the asymmetry observed through JWST is stronger than the magnitude of the asymmetry reported previously using Earth-based data. That can be linked to the previous observation that the magnitude of the asymmetry increases as the redshift gets higher \citep{shamir2019large,shamir2020patterns,shamir2022large,sym16101389}. For instance, Table~\ref{dis_120_210}, taken from \citep{shamir2020patterns}, shows the distribution of spiral galaxies imaged by SDSS at different redshift ranges. The RA is limited to $(120^o,210^o)$, which is around the location of the Northern end of the Galactic pole.

\begin{table}
{
\begin{tabular}{|l|c|c|c|c|}
\hline
z &    cw    &  ccw  & ${cw}\over{cw+ccw}$ & p-value \\      
\hline
0-0.05     &	3216 &	 3180 & 	0.5003	& 0.698  \\
0.05-0.1 &	6240	& 6270 &	0.498 &	0.4 \\
0.1-0.15 &	4236 &	4273 & 	0.496 &	0.285  \\
0.15-0.2 &	1586 &	1716 & 	0.479 &	0.008  \\
0.2 - 0.5 &	2598 &	 2952 &	 0.469 &	$1.07\cdot10^{-6}$  \\

\hline
Total & 17,876 & 18,391 & 0.493 & 0.0034 \\
\hline
\end{tabular}
\caption{The distribution of galaxies rotating clockwise and counterclockwise imaged by SDSS. All galaxies are within the RA range of $(120^o,210^o)$. The p-values are the binomial distribution p-value to have such asymmetry or stronger by chance. The table is taken from \citep{shamir2020patterns}.}
\label{dis_120_210}
}
\end{table}

As the table shows, the asymmetry increases as the redshift gets higher. While there is no proven link between the observation made with JWST and the information provided in Table~\ref{dis_120_210}, it should remain a possibility that the observations are linked. In that case, the asymmetry changes gradually with time or distance from Earth.

Possible explanations to the observation can  be broadly divided into two categories: The first is an anomaly in the large-scale structure (LSS) of the early universe, and the second is related to the physics of galaxy rotation.  

\subsection{Anomaly in the LSS}

If the observation shown here indeed reflects the structure of the Universe, it shows that the early universe was more homogeneous in terms of the directions towards which galaxies rotate, and becomes more chaotic over time while exhibiting a cosmological-scale axis that is close to the Galactic pole. Some cosmological models assume a geometry that features a cosmological-scale axis. These include ellipsoidal Universe 
\citep{campanelli2006ellipsoidal,campanelli2007cosmic,campanelli2011cosmic,gruppuso2007complete,cea2014ellipsoidal}, dipole big bang \citep{allahyari2023big,krishnan2023dipole}, and isotropic inflation 
\citep{arciniega2020geometric,edelstein2020aspects,arciniega2020towards,jaime2021viability,feng2003double,piao2004suppressing,bohmer2008cmb,luongo2022larger,dainotti2022evolution}. In these cases, the large-scale distribution of the galaxy rotation is aligned in the form of a cosmological-scale axis, and the location of that axis in close proximity to the Galactic pole can be considered a coincidence.

An additional cosmological model that requires the assumption of a cosmological-scale axis is the theory of rotating Universe \citep{godel1949example,ozsvath1962finite,ozsvath2001approaches,sivaram2012primordial,chechin2016rotation,seshavatharam2020integrated,camp2021}. That model is also related to the theory of black hole cosmology
\citep{pathria1972universe,stuckey1994observable,easson2001universe,chakrabarty2020toy,tatum2018flat}, according which the Universe is the interior of black hole in a parent universe, and therefore is also aligned with the contention of multiverse.

Because black holes spin
\citep{mcclintock2006spin,mudambi2020estimation,reynolds2021observational}, a universe hosted inside of a black hole is also expected to spin. Therefore, it has been proposed that a universe located in the interior of a black hole should have an axis, and inherit the preferred direction of the host black hole
  \citep{poplawski2010cosmology,seshavatharam2010physics,christillin2014machian,seshavatharam2020light,seshavatharam2020integrated}.
Black hole cosmology is also linked to the theory of holographic universe \citep{susskind1995world,bak2000holographic,bousso2002holographic,myung2005holographic,hu2006interacting,rinaldi2022matrix,sivaram2013holography,shor2021representation}.

Another paradigm relevant to the observation described here is the contention that the large-scale structure of the Universe has a fractal structure, reflected by fractal patterns formed by the large-scale distribution galaxies  \citep{gabrielli2005fractals,teles2022galaxy,baryshev2000conceptual,labini2001complexity,labini2004complexity,baryshev1998fractal,pietronero2000fractal,labini2003fractals}.  These patterns challenge the assumption that the distribution of galaxies in the large-scale structure is random.
  
These explanations are considered alternative to the standard cosmological model \citep{turner1996cosmology,pecker1997some,perivolaropoulos2014large,bull2016beyond,velten2020hubble,netchitailo2020world}, and also violates the isotropy assumption of the Cosmological Principle. Although the Cosmological Principle is the basic assumption for the standard cosmological model, its correctness has been challenged \citep{pecker1997some,kroupa2012dark}. Observations using a variety of different probes have also challenged the correctness of the Cosmological Principle in an empirical manner \citep{Aluri_2023}.

\subsection{Physics of galaxy rotation}

As mentioned above, if the distribution of directions of rotation of galaxies indeed form a cosmological-scale axis, the alignment of that axis with the Galactic pole could be considered a coincidence. But another explanation could be that the distribution of galaxy direction of rotation in the Universe is random, but only seems non-random to an Earth-based observer. In that case, the observation can be explained by the effect of the rotational velocity of the observed galaxies relative to the rotational velocity of Earth around the center of the Milky Way galaxy. That can explain the observation without violating the Cosmological Principle. The proximity to the Galactic pole is expected, as the difference between the rotational velocity of the Milky Way and the rotational velocity of the observed galaxies peaks at the Galactic pole.

As discussed in \citep{shamir2017large,shamir2020asymmetry,sym15061190}, due to the Doppler shift effect galaxies that rotate in the opposite direction relative to the Milky Way are expected to be slightly brighter than galaxies that rotate in the same direction relative to the Milky Way. Therefore, more galaxies that rotate in the opposite direction relative to the Milky Way are expected to be observed from Earth, and the difference should peak at around the Galactic pole. That observation is conceptually aligned with the empirical data of Figure~\ref{jwst_location}, and the observation using JADES described in Section~\ref{results}.

This explanation is challenged by the fact that the effect of the rotational velocity have merely a mild impact on the brightness of galaxies, and therefore is not expected to lead to the dramatic difference of $\sim$50\% in the number of galaxies as observed through JADES. On the other hand, empirical observations showed that the difference in brightness is larger than expected given the rotational velocity of galaxies \citep{shamir2017large,shamir2020asymmetry,sym15061190}. That was observed with SDSS \citep{sym15061190}, Pan-STARRS \citep{shamir2017large}, and the space-based HST \citep{shamir2020asymmetry}. Similar observations were made with the redshift \citep{shamir2024simple,particles7030041}, also showing that the magnitude of the asymmetry increases as the redshift gets larger \citep{particles7030041}. Other related observations can be the dipole formed by quasar distribution as observed from Earth \citep{hutsemekers2014alignment,quasars}, which has been shown to be linked to the colour \citep{panwar2024colour}, and therefore could also be a photometric effect rather than a feature of the large-scale structure of the Universe. The number of galaxies in the line-of-sight also show a surprising cosmological-scale anisotropy \citep{ahn2025kinematically}, and that can also be related to the differences in the brightness of galaxies as observed from Earth.

The difference can be linked to the mysterious physics of galaxy rotation, which is known to be in substantial tension with the mass \citep{oort1940some,rubin1983rotation}. Common explanations include dark matter \citep{rubin1983rotation,el2020aedge}, modified Newtonian dynamics \citep{milgrom1983modification}, and others \citep{sanders1990mass,capozziello2012dark,chadwick2013gravitational,farnes2018unifying,rivera2020alternative,nagao2020galactic,blake2021relativistic,gomel2021effects,skordis2021new,larin2022towards}, but no explanation has been fully proven. In particular, the theory of dark matter as the explanation to the difference between the mass and rotational velocity of stars within galaxies has been challenged, and despite over a century of research there is still no clear proven explanation to the physics of galaxy rotation \citep{sanders1990mass,mannheim2006alternatives,kroupa2012dark,kroupa2012failures,kroupa2015galaxies,arun2017dark,akerib2017,bertone2018new,aprile2018,skordis2019gravitational,sivaram2020mond,hofmeister2020debate,byrd2021spiral}. Therefore, it is possible that the physics of galaxy rotation, which is not yet fully known, affects the brightness of the galaxy in a manner that is not necessarily expected.


If the physics of galaxy rotation affects the light that galaxies emit in a manner that is currently unknown, that can also affect the redshift, and therefore can be related to alternative redshift models  \citep{crawford1999curvature,pletchermature,lovyagin2022cosmological,seshavatharam2023rotating,unknown,gupta2401testing,lee2023cosmological,shao2013energy,kragh2017universe,shao2018tired,sato2019tired,laviolette2021expanding,lovyagin2022cosmological,lopez2023history}.  Although the physical mechanism of such phenomenon is not clear, using alternative redshift models can explain a large number of observations that are currently unexplained such as dark energy, the $H_o$ tension \citep{wu2017sample,mortsell2018does,bolejko2018emerging,davis2019can,pandey2020model,camarena2020local,di2021realm,riess2022comprehensive}, as well as the unexpected presence of large and massive galaxies in the early Universe \citep{xiao2023massive,glazebrook2024massive} that challenge the age of the Universe as estimated by the existing models. The age of the Universe has been challenged also by the presence of stars that are older than the estimated age of the Universe such as HD 140283 \citep{guillaume2024age}. These tensions challenge modern cosmology, and trigger a variety of solutions and explanations that involve new physics. These puzzling observations can be solved by using an alternative the redshift model \citep{crawford1999curvature,pletchermature,lovyagin2022cosmological,seshavatharam2023rotating,unknown,gupta2401testing,gupta2023jwst,lee2023cosmological,shao2013energy,kragh2017universe,shao2018tired,sato2019tired,laviolette2021expanding,lovyagin2022cosmological,lopez2023history}.   Although the physics that can lead to alternative redshift models is also not yet known, it can explain the observed tensions regarding the expansion rate and age of the Universe. 

The unprecedented power of JWST, combined with other recent observations have revolutionized cosmology, and triggered substantial changes in the studying of the Universe. It is likely that research efforts to explain them will continue in the next few decades. The observation reported here can provide yet another piece of information that can ultimately lead to a complete model that can provide a consolidated explanation to all current unexplained observations.

\section*{Data Availability}

The data used in this study is available in the tables included in the manuscript. URLs to relevant datasets from previous papers are also provided in the body of the manuscript. 


\section*{Acknowledgments}

I would like to thank the knowledgeable reviewer and associate editor for the insightful comments. The study was supported in part by National Science Foundation grant 2148878.


\bibliographystyle{apalike}
\bibliography{main}

\bsp	
\label{lastpage}
\end{document}